\documentclass[10pt,pra,twocolumn,floatfix]{revtex4}
\usepackage{amssymb}
\usepackage{amsfonts}
\usepackage{amssymb}
\usepackage{amsmath}
\usepackage[dvips]{graphicx}
\usepackage{color}
\usepackage{appendix}

\usepackage{graphicx}
\usepackage{dcolumn}
\usepackage{bm}

\usepackage{graphicx}

\usepackage{hyperref}

\usepackage{mathrsfs}
\usepackage{isomath}
\usepackage{amsthm}
\usepackage{epstopdf}
\usepackage{txfonts}

\begin{document}

\title{Remote preparation and manipulation of squeezed light}

\author{Dongmei Han$^{1,2}$, Na Wang$^{1,2}$, Meihong Wang$^{1,2}$, Zhongzhong Qin$^{1,2}$, Xiaolong Su$^{1,2}$}

\email{suxl@sxu.edu.cn}

\affiliation{$^{1}$State Key Laboratory of Quantum Optics and Quantum Optics Devices, \\
Institute of Opto-Electronics, Shanxi University, Taiyuan, 030006, People's
Republic of China \\
$^{2}$Collaborative Innovation Center of Extreme Optics, Shanxi University,\\
Taiyuan, Shanxi 030006, People's Republic of China\\
}

\begin{abstract}
Remote state preparation enables one to create and manipulate a quantum state based on the shared entanglement between distant nodes. 
Here, we experimentally demonstrate remote preparation and manipulation of squeezed light. By performing homodyne projective measurement on one mode of the continuous variable entangled state at Alice's station, a squeezed state is created at Bob's station. Moreover, rotation and displacement operations are applied on the prepared squeezed state by changing the projective parameters on Alice's state. We also show that the remotely prepared squeezed state is robust against loss and $N-1$ squeezed states can be remotely prepared based on a $N$-mode continuous variable Greenberger-Horne-Zeilinger-like state. Our results verify the entanglement-based model used in security analysis of quantum key distribution with continuous variables and have potential application in remote quantum information processing.  
\end{abstract}

\maketitle

\section{Introduction}

With the development of the quantum network, it becomes possible for users who do not have the ability of preparing quantum state to obtain a quantum resource and implement quantum information processing. Remote state preparation (RSP) enables one to create and manipulate quantum states remotely based on shared entanglement \cite{Lo2000,Paris2003continuous}. Compared with direct state transmission, where a prepared quantum state is transmitted to the user through a quantum channel, RSP offers remote control of the quantum state and intrinsic security \cite{Pogorzalek2019}. Compared with quantum teleportation, RSP does not need joint measurement, requires fewer classical communication \cite{Aru2000} and offers the ability to manipulate quantum state remotely. Based on RSP protocol, the single photon state \cite{Lvovsky2001}, sub-Poissonian state \cite{Laurat2003}, superposition state up to two-photon level \cite{Erwan2010}, cat state \cite{Ulanov2016}, continuous variable qubits \cite{Laurat2018} and squeezed state in the microwave regime \cite{Pogorzalek2019} have been experimentally demonstrated. 

A squeezed state has broad applications in continuous variable (CV) quantum information \cite{Lvovsky2015,Braunstein2005,Weedbrook2012,wangxb2007,su2020}, quantum measurement \cite{LIGO2013,Schnabel2020} and quantum-enhanced imaging \cite{michael2013,catxere2021}. Up to now, a squeezed state is prepared locally based on an optical parametric amplifier \cite{Wu1986,Takeno2007,McKenzie2004,su2016,Henning2008}, four-wave mixing \cite{Slusher1985,Guo2014,Corzo2011} and a photonic chip \cite{Zijiao2021,Zhang2021}. In addition to the local preparation, it has been proposed that a squeezed state can also be prepared based on the RSP protocol by performing homodyne measurement on one mode of a CV Einstein-Podolsky-Rosen (EPR) entangle state \cite{Paris2003continuous}. This RSP protocol corresponds to the entanglement-based model widely used in the security analysis of CV quantum key distribution (QKD) \cite{Frederic2003,Ibl,Gros2,Nav1}, where the security of CV QKD is analyzed based on a CV EPR entangled state since it has been shown that CV QKD with squeezed state (coherent state) is equivalent to homodyning (heterodyning) one mode of a EPR entangled state \cite{Frederic2003}. However, remote preparation and manipulation of squeezed states by homodyne projective measurement have not been experimentally demonstrated.

Here, we experimentally demonstrate the remote preparation of squeezed states based on a CV EPR entangled state distributed between Alice and Bob.  By performing homodyne projective measurement on Alice's state, a squeezed state with approximately $-$1.27 dB squeezing and fidelity of 92\% is remotely prepared at Bob's station. Then, the prepared squeezed state is rotated and displaced by changing the parameters of the homodyne projective measurement at Alice's station, which is equivalent to performing rotation and displacement operations on the squeezed state. We show that the remotely prepared squeezed state is robust against loss in the quantum channel. Furthermore, this scheme is extended to RSP based on an $N$-mode CV Greenberger-Horne-Zeilinger-like (GHZ-like) state $\int dx |x,x,x,...,x \rangle$, which is the eigenstate with total momentum (phase quadrature) zero $p_1+p_2+p_3+... +p_N = 0$ and relative positions (amplitude quadratures) $x_i - x_j = 0$ ($i, j = 1, 2, 3,...N$) \cite{van2000,loock2002}, where $N-1$ squeezed states can be remotely prepared simultaneously by performing homodyne projective measurement on one mode of the CV GHZ-like state. The presented results provide a new method to prepare and manipulate squeezed states remotely.

\section{Remote preparation scheme}

\begin{figure}[tbp]
\begin{center}
\includegraphics[width=\linewidth]{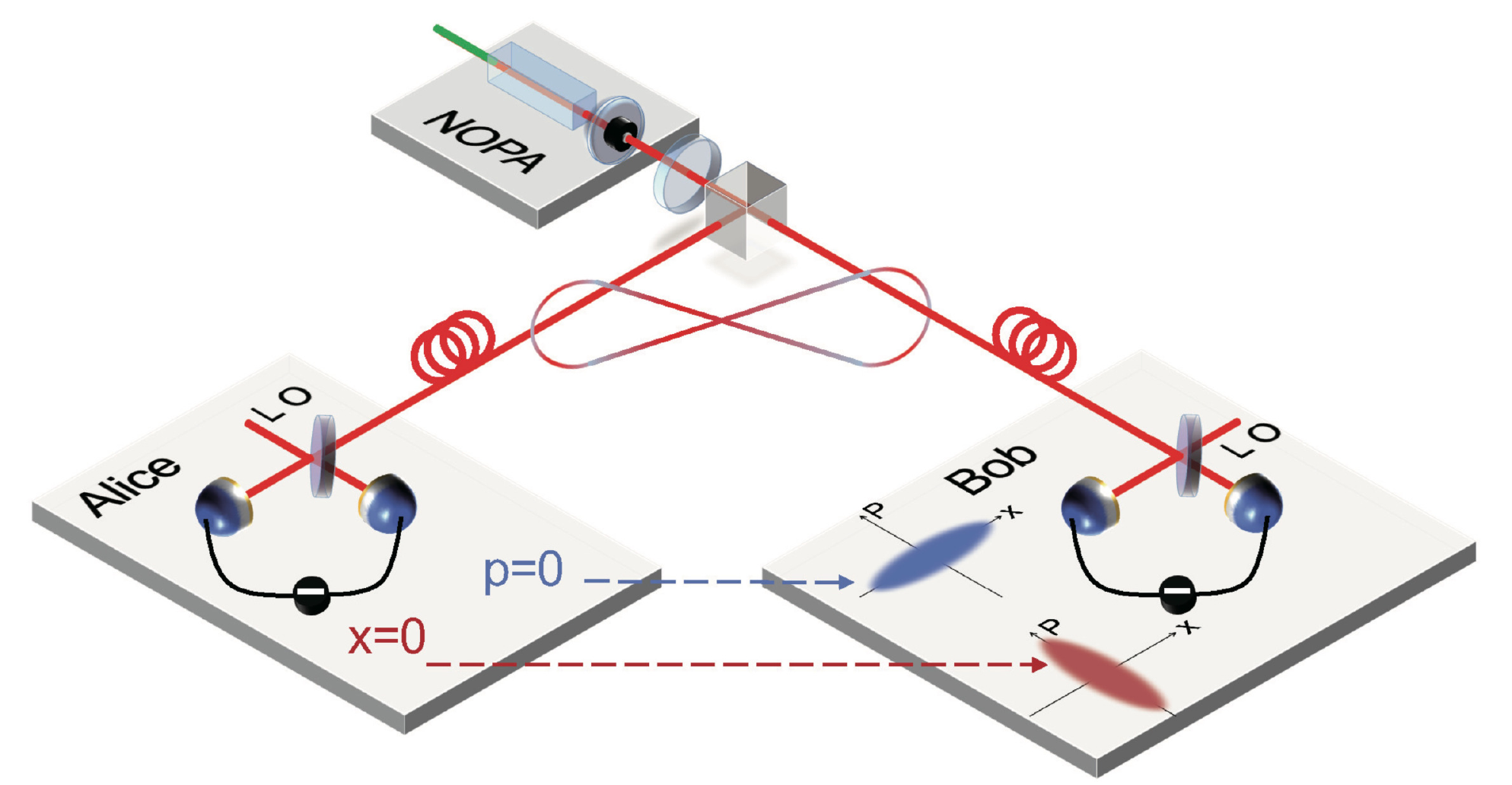}
\end{center}
\caption{Experimental set-up. Two modes of a CV EPR entangled state are separated by a polarization beam splitter (PBS) and distributed to Alice and Bob respectively. Alice performs a homodyne projective measurement on her mode. The  phase-squeezed state or amplitude-squeezed state is prepared at Bob's station conditioned on the measurement results of $p_{A} = 0$ or $x_{A} = 0$. A lossy channel is simulated by the combination of a half-wave plate and a PBS. LO, Local oscillator.}
\label{Fig1}
\end{figure}

As shown in Fig.~\ref{Fig1}, a CV EPR entangled state is prepared by a non-degenerate optical parametric amplifier (NOPA) in a quantum server and distributed to Alice and Bob through two lossy channels. By measuring the phase quadrature of Alice's state and projecting the quadrature values to $p_{A}=0$, a phase squeezed state is prepared at Bob's station remotely. By measuring the amplitude quadrature of Alice's state and projecting the quadrature values to $x_{A}=0$, the squeezed state is rotated by 90 degrees, i.e. an amplitude squeezed state is prepared at Bob's station. The projective measurement is implemented by the post-selection of quadrature values with a selection width of $|\delta x| < 0.1$. By projecting the quadrature values to $x_{A}=\alpha$ ($p_{A}=\alpha$), the displacement operation can be applied on the remotely prepared squeezed states. 

A CV EPR entangled state can be fully characterized by its covariance matrix which is expressed by
\begin{equation}
\sigma_{A B}
=\left(\begin{array}{cccc}
\Delta^{2}\hat{x}_{A}& 0 & \Delta^{2}(\hat{x}_A\hat{x}_B) & 0 \\ 
0 & \Delta^{2}\hat{p}_{A} & 0 & \Delta^{2}(\hat{p}_A\hat{p}_B)  \\ 
\Delta^{2}(\hat{x}_B\hat{x}_A)  & 0 & \Delta^{2}\hat{x}_{B} & 0 \\ 
0 & \Delta^{2}(\hat{p}_B\hat{p}_A)  & 0 & \Delta^{2}\hat{p}_{B} 
\end{array}\right),
\end{equation}
where $\hat{x}_{A(B)}=(\hat{a}_{A(B)}^{\dagger}+\hat{a}_{A(B)})/\sqrt{2}$ and $\hat{p}_{A(B)}=i(\hat{a}_{A(B)}^{\dagger}-\hat{a}_{A(B)})/\sqrt{2}$ denote amplitude and phase operators where $\hat{a}^{\dagger}, \hat{a}$ are creation and annihilation operators respectively. We have $\Delta^{2}\hat{x}_{A}$ = $\Delta^{2}\hat{p}_{A}$ = $[\eta_{A}({V_a+V_s}) + (1-\eta_{A})]/2$, $\Delta^{2}\hat{x}_{B}$ = $\Delta^{2}\hat{p}_{B}$ = $[\eta_{B}(V_a+V_s) + (1-\eta_{B})]/2$, and $\Delta^{2}(\hat{x}_{A}\hat{x}_{B})$ = $-\sqrt{\eta_{A} \eta_{B}}(V_s-V_a)/2$,  $\Delta^{2}(\hat{p}_{A}\hat{p}_{B})$ = $\sqrt{\eta_{A} \eta_{B}}(V_s-V_a)/2$. Here, $V_s$ and $V_a$ represent variances of squeezed and anti-squeezed quadratures, respectively. We have $V_a  V_s = 1/4$ for a pure state and $V_a  V_s > 1/4$ for a mixed state. Additionally, $\eta_A$ and $\eta_B$ represent transmission efficiencies of Alice's and Bob's modes respectively. The corresponding Wigner function of the CV EPR entangled state is given by \cite{Weedbrook2012}
\begin{equation}
\begin{aligned}
W_{AB}(x_{A},p_{A},x_{B},p_{B})=\frac{1}{\sqrt{Det\sigma_{AB}}{\pi}^{2}}\exp \left\{-\frac{1}{2}\left(\xi^{\top}\sigma_{AB}^{-1} \xi \right)\right\},
\end{aligned}
\end{equation}
where $\hat{\xi}\equiv (\hat{x}_{A},\hat{p}_{A},\hat{x}_{B},\hat{p}_{B})^{\top}$ is the vector of the amplitude and phase quadratures of the entangled state. The Wigner function of the homodyne projective measurement $\Pi_x$ on amplitude quadrature of Alice's state is expressed by \cite{Paris2003continuous}
\begin{equation}
\begin{aligned}
W[\Pi_{x}](x_{A})=\delta(x_{A} - \alpha),
\end{aligned}
\end{equation}
where $\alpha$ is the projective value. After the Alice's homodyne projective measurement, Bob's state is collapsed to
\begin{equation}
\begin{aligned}
W_B(x_{B}, p_{B})=\int\int dx_{A}dp_{A}W_{AB}\times W[\Pi_{x}](x_{A}).
\end{aligned}
\end{equation}
For example, if Alice chooses $x_A = 0$ ($\alpha = 0$), Bob's state becomes $W_B(x_{B}, p_{B})=\int dp_{A}W_{AB}(0,p_{A},x_{B},p_{B})$ which has unit fidelity with an ideal amplitude-squeezed state if the entangled state is pure. Similarly, homodyne projective measurement $\Pi_p$ on phase quadrature projects Bob's mode to a phase-squeezed state. Furthermore, for a CV EPR entangled state with correlated amplitude and anti-correlated phase quadratures in the case of infinite squeezing, the projective measurement of $x_A = \alpha$ leads to an amplitude squeezed state displaced by ($\alpha$, 0) in phase space at Bob's station \cite{Frederic2003}. Similarly, by projecting on $p_A = \alpha$, a phase squeezed state displaced by (0, $-\alpha$) can be prepared.

\begin{figure}[tbp]
\begin{center}
\includegraphics[width=\linewidth]{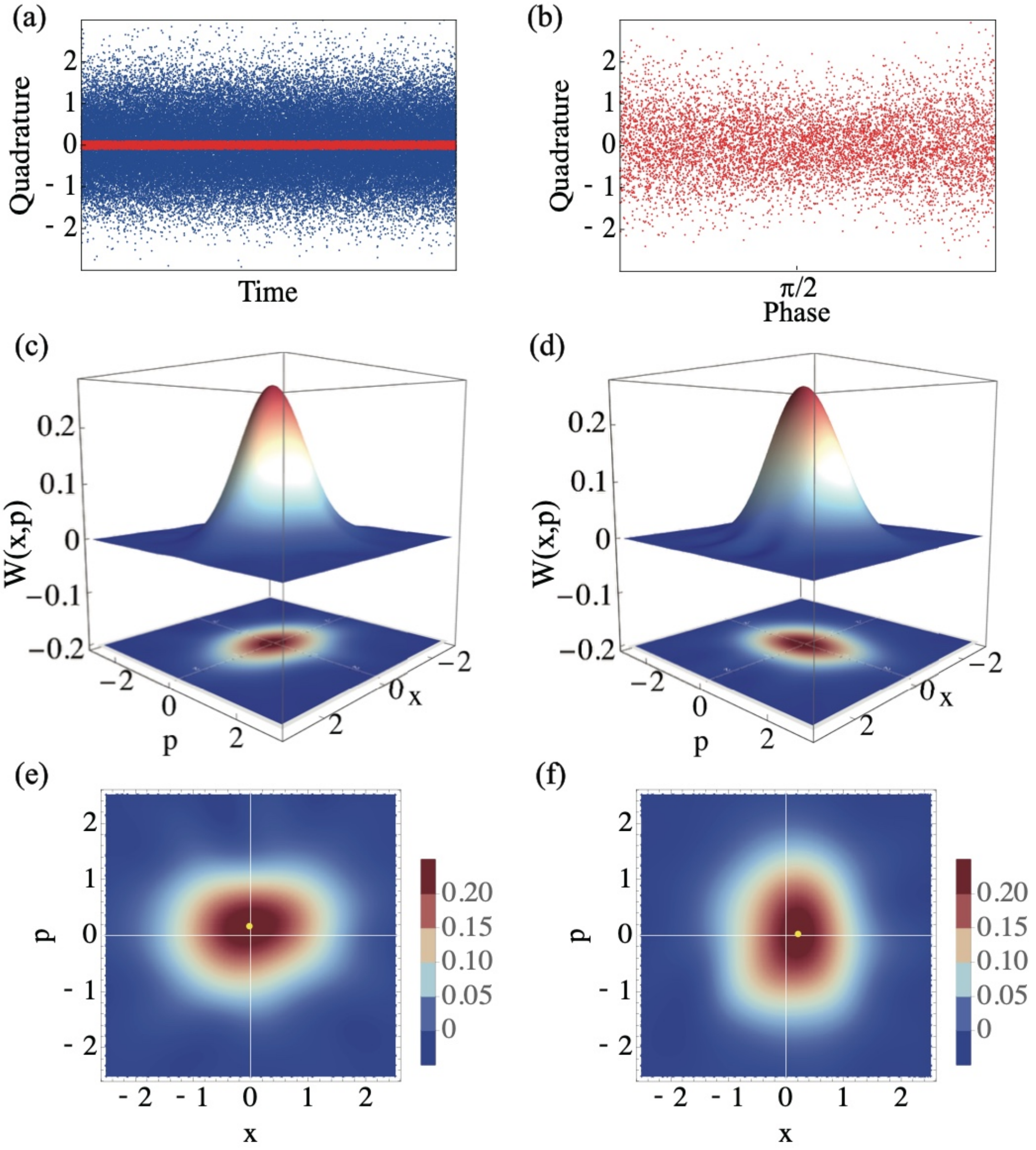}
\end{center}
\caption{
\textbf{(a),} Quadrature values of Alice's mode (70000 and 7000 for blue and red data points).\textbf{(b),} Quadrature values of Bob's mode (7000 data points).\textbf{(c,d),} Reconstructed Wigner functions of phase and amplitude squeezed states. \textbf{(e,f),} Contour plots of phase and amplitude-squeezed states displaced by (-0.01,0.17) and (0.20,0.05) respectively in phase space. Around 10000 data points each are used to reconstruct these Wigner functions.}
\label{Fig2}
\end{figure}

\section{Experimental Setup and Results}

In our experiment, the NOPA is composed by a 10-mm-long $\alpha$-cut type-II potassium titanyl phosphate (KTP) crystal and a concave mirror with 50-mm radius. The front face of the KTP crystal is coated to be used for the input coupler and the concave mirror serves as the output coupler of the NOPA. The details of parameters of the NOPA have been provided elsewhere \cite{liu2019,deng2021,wang2020}. Our NOPA cavity is locked by using the Lock-and-hold technique (See Appendix A for details). The seed beam is injected into the NOPA for the cavity locking during the locking period, while it is chopped off to perform the measurement during the hold period. The NOPA works at amplification status, where the relative phase between seed and pump beam is locked to 0, a CV EPR entangled state with correlated amplitude ($\langle\Delta^2(\hat x_A - \hat x_B)\rangle=e^{-2r}$) and anti-correlated phase quadratures ($\langle\Delta^2(\hat p_A + \hat p_B)\rangle=e^{-2r}$) is obtained, where $r$ is the squeezing parameter. a CV EPR entangled state with $V_{s} = 0.24$ (corresponding to $-$3.2 dB squeezing) and $V_{a} = 1.3$ (corresponding to 4.2 dB antisqueezing) in the bandwidth of 60 MHz is prepared when the NOPA is pumped at 70 mW. We lock the relative phase between the signal and local oscillator of the homodyne detector (HD) at Alice's station to 0 and 90 degrees to measure the amplitude quadrature $\hat x_A$ and phase quadrature $\hat p_A$, respectively. The output signals of two HDs are filtered by two 60 MHz low-pass filters respectively and recorded simultaneously by a digital storage oscilloscope. Bob performs quantum tomography to reconstruct the Wigner function of his state.

\begin{figure}[tbp]
\includegraphics[width=\linewidth]{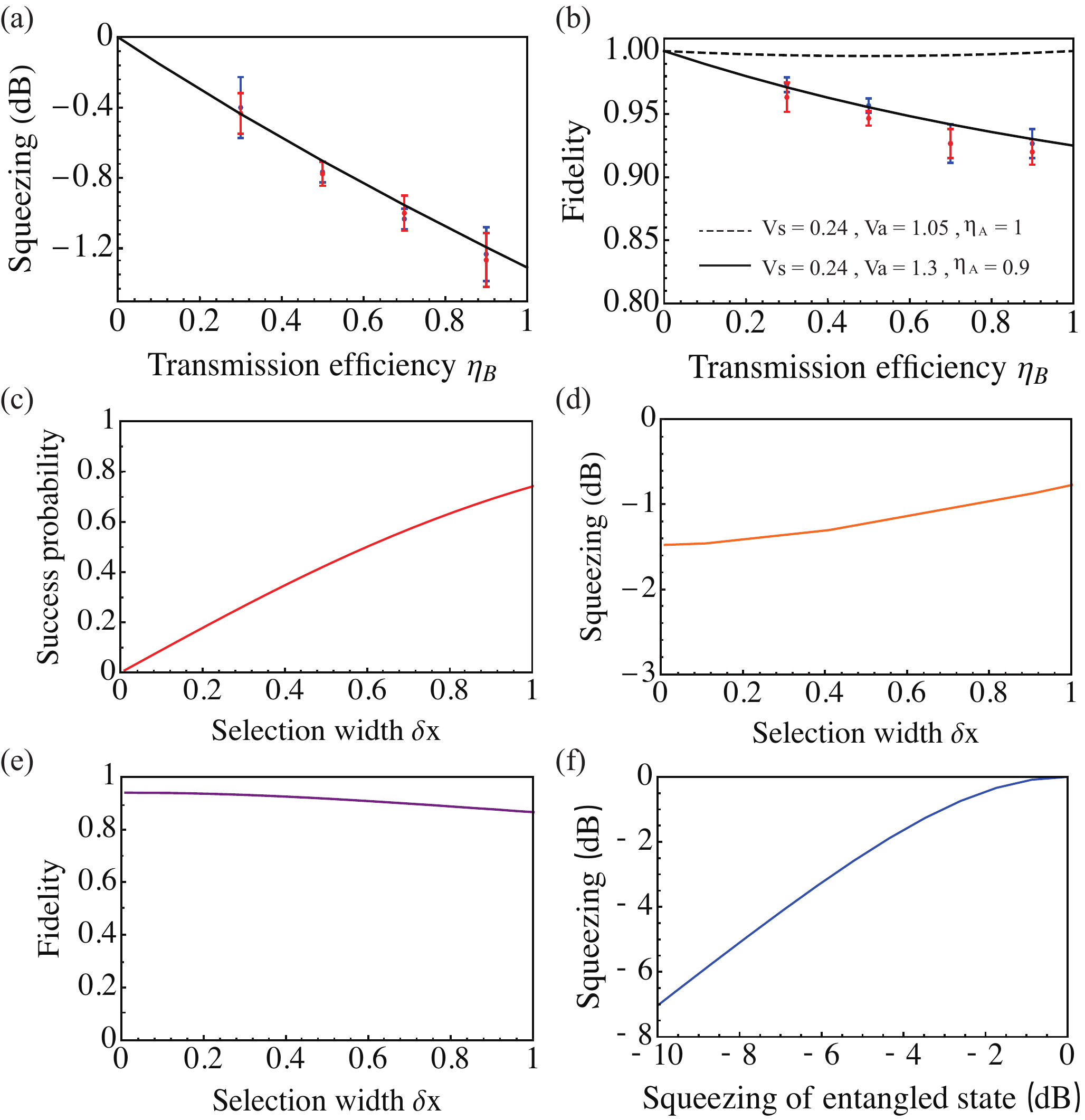}
\caption{
\textbf{(a,b),} Squeezing level and fidelity of the prepared squeezed states as functions of the transmission efficiency of Bob's mode. The red and blue data points represent phase squeezed and amplitude squeezed states respectively and the error bars are obtained by standard deviation of measurements repeated three times. \textbf{(c - e),} Dependence of success probability, squeezing level and fidelity of the prepared squeezed state on the selection width $\delta x$. \textbf{(f),} The dependence of squeezing level of the prepared squeezed state on squeezing level of a pure entangled state.}
\label{Fig3}
\end{figure}

The measured phase quadratures of Alice's mode (blue data points) in the time domain are shown in Fig.~\ref{Fig2}(a), where the red points represent the data satisfying the condition $|\delta x|< 0.1$. Conditioned on the red data points of Alice's quadrature values, the selected data points of Bob's mode are shown as the red points in Fig.~\ref{Fig2}(b). As shown in Fig.~\ref{Fig2}(c), a phase squeezed-state with around $-$1.27 dB squeezing is remotely prepared at Bob's station by choosing $p_A = 0$ when the transmission efficiency between Alice and Bob is 81\% ($\eta_A =\eta_B = 0.9$). By changing the projective basis to $x_A = 0$, an amplitude-squeezed state with squeezing level around $-$1.26 dB is remotely prepared too, as shown in Fig.~\ref{Fig2}(d). This is equivalent to applying a rotation operation of 90 degrees in phase space, which turns a phase-squeezed state into an amplitude-squeezed state. In principle, rotation operation with arbitrary degree can be implemented by locking the relative phase of the homodyne detector to arbitrary degree (See Appendix B for details). 

We quantify the quality of the remotely prepared squeezed state by the fidelity $F=\int\int dxdp W^{x(p)}_{ds}W(x,p)$ which is defined as the overlap between the experimentally prepared state $W(x,p)$ and theoretically predicted state $W^{x}_{ds} = \frac{1}{\pi} \exp(-\frac{(x-a)^{2}}{e^{-2r}}-\frac{(p-b)^{2}}{e^{2r}})$ ($W^{p}_{ds} = \frac{1}{\pi} \exp(-\frac{(x-a)^{2}}{e^{2r}}-\frac{(p-b)^{2}}{e^{-2r}})$), where $a$ and $b$ represent the displacements along the amplitude and phase quadratures, respectively.
The estimated squeezing parameter of the state corresponds to the value where the fidelity riches its maximum. The fidelities of the prepared squeezed states with a transmission efficiency of 81\% are all above 92\%.

The remotely prepared squeezed states are displaced in phase space by changing the projective measurement to $p_A\approx $ $-$0.4 and $x_A\approx $ 0.4, as shown in Figs.~\ref{Fig2}(e) and~\ref{Fig2}(f). Squeezed states with $-$1.2 dB squeezing and fidelities of 83\% and 81\% are displaced, respectively. The displacement in the amplitude (phase) direction of the amplitude (phase) squeezed state is limited by the squeezing level of the entangled state (See Appendix B for details). These results demonstrates the validity of the entanglement-based model used in security analysis of CV QKD.

To verify the dependence of remotely prepared state on channel loss, Bob's mode is transmitted through a lossy channel when $\eta_A = 0.9$. As shown in Fig.~\ref{Fig3}(a), it is obvious that the squeezing of the remotely prepared squeezed state is robust against loss in the quantum channel, which has the same robustness as that of directly transmitting a squeezed state in a lossy channel~\cite{deng2016}. The fidelity of the prepared state decreases a little bit with the increase of transmission efficiency [Fig.~\ref{Fig3}(b)], which is because that our initial entangled state is not pure and 10\% loss exists in Alice's channel. If there is no loss in Alice's channel and the entangled state is pure, the fidelities of the remotely prepared squeezed states are always near unit after the quantum channel [dotted curve in Fig.~\ref{Fig3}(b)].

In our experiment, the success probability of the remotely prepared squeezed state is $10\%$ corresponding to the selection width $|\delta x| < 0.1$. Here, we analyze the effect of the selection width $\delta x$ on the prepared squeezed state with the transmission efficiencies $\eta_A = \eta_B = 1$. Conditioned on the homodyne projective measurement, the remotely prepared state is given by $W_B(x_{B}, p_{B})=\int^{x+\delta x}_{x-\delta x} \int dx_{A}dp_{A}W_{AB}(x_{A},p_{A},x_{B},p_{B})$ when the measured quadratures are selected within a range $x_A \in [x-\delta x , x+\delta x]$. It is obvious that remotely prepared state is related to the selection width $\delta x$. The success probability is increased with the increase of selection width, while the fidelity and squeezing of the prepared state are decreased [Fig.~\ref{Fig3}(c-e)]. Thus it is necessary to chose a suitable selection width to balance the trade-off between success probability and fidelity. The squeezing of the remotely prepared squeezed state can be improved by increasing the squeezing of the entangled state, as shown in Fig.~\ref{Fig3}(f). 

\begin{figure}[tbp]
\includegraphics[width=\linewidth]{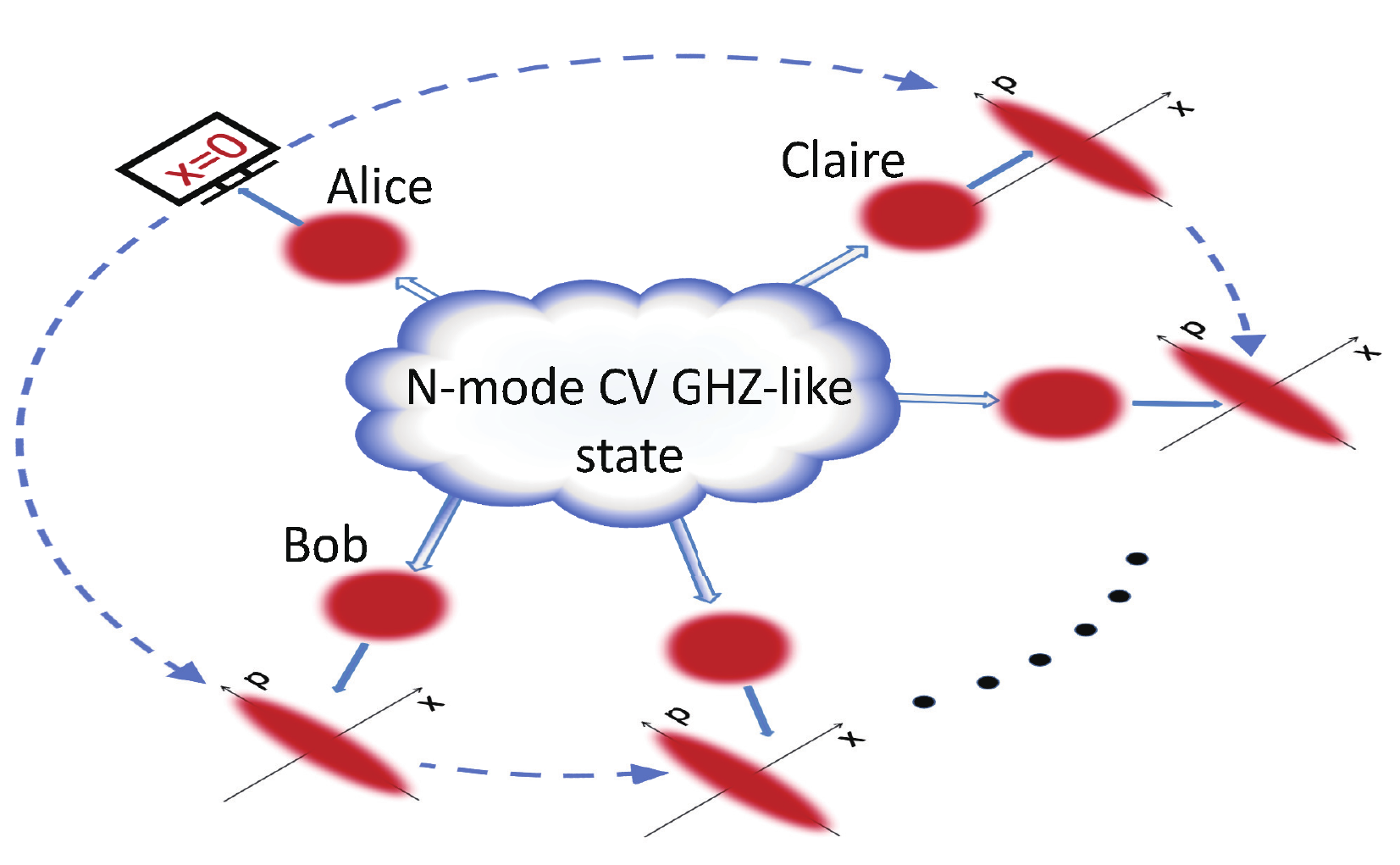}
\caption{ Remote preparation of $N$-1 squeezed states based on an N-mode multipartite CV GHZ-like state. By performing amplitude quadrature projective measurement on an arbitrary mode, for example $x_A = 0$, the other modes collapse to amplitude squeezed states.}
\label{Fig4}
\end{figure}

We also extend this RSP protocol to a $N$-mode CV GHZ-like state, which enables users in a quantum network to obtain squeezed states. A quantum network is established by distributing optical modes of a CV GHZ-like state to spatially separated quantum nodes, as shown in Fig.~\ref{Fig4}. The wave function of amplitude quadrature is given by $\psi_{N}(x) = (\frac{1}{\pi})^{\frac{N}{4}}\exp[-e^{-2r}\frac{1}{2N}(\sum^{N}_{i=1} x_i)^2-e^{2r}\frac{1}{4N}\sum^{N}_{i,j}(x_i-x_j)^2]$, which is proportional to $\psi_{N}(x) \propto \delta(x_i - x_j)$ in the limit of infinite squeezing \cite{loock2002}. If we project one mode to $x_i = 0$, the rest modes will be proportional to $\psi_j(x) \propto \delta(x_j)$, i.e. amplitude-squeezed states can be created at $N$-1 remote stations simultaneously. If the user in the network needs a p-squeezed state, it requires $N$-1 projective measurements collaboratively performed by corresponding users. More details about the squeezed state preparation from the multimode CV GHZ-like state can be found in Appendix C.
After the RSP based on the CV GHZ-like state, the users can further implement quantum information processing based on a squeezed state at their own station respectively.

\section{Conclusion}

In summary, we prepare squeezed states remotely based on the shared CV EPR entangled state by performing homodyne projective measurement at Alice's station. The prepared squeezed state is rotated and displaced by changing the parameters of homodyne detection. More importantly, we show that the prepared squeezed state is robust against loss in the quantum channel. We also show that this RSP scheme can be extended to the multipartite CV entangled state, where $N$-1 squeezed states are prepared simultaneously by performing homodyne projective measurement at one station based on a shared $N$-mode CV GHZ-like state. Our results present the evidence for the entanglement-based model used in security analysis of CV QKD and make a crucial step toward the remote quantum information processing.

\section{ACKNOWLEDGMENTS}

This research was supported by the NSFC (Grants Nos. 11834010, 11974227, and 62005149), and the Fund for Shanxi \textquotedblleft 1331 Project\textquotedblright\ Key Subjects Construction.

\section*{APPENDIX A: DETAILS OF THE EXPERIMENT}

\begin{figure}[h]
\setcounter{figure}{0}
\renewcommand*{\thefigure}{A\arabic{figure}}
\begin{center}
\includegraphics[width=0.85\linewidth]{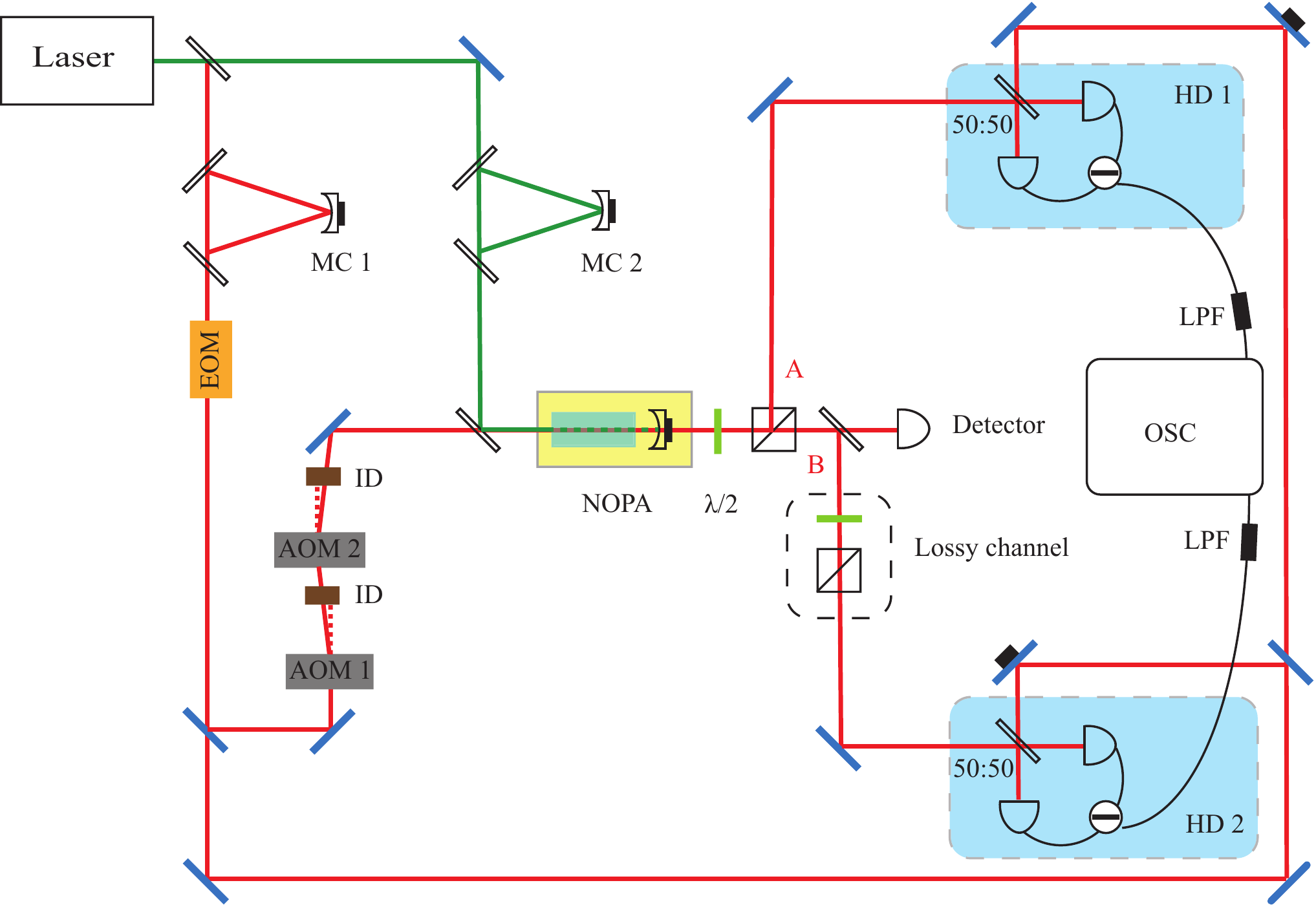}
\end{center}
\caption{\textbf{Experimental setup.} MC 1: mode cleaner for the $1080$ nm light, MC 2: mode cleaner for the $540$ nm light, EOM: electro-optic modulator, AOM: acousto-optic modulator, ID: iris diaphragm, NOPA: non-degenerate optical parametric amplifier, LPF: low-pass filter, HD 1: homodyne detector at Alice's station, HD 2: homodyne detector at Bob's station, OSC: oscilloscope.}
\label{A1}
\end{figure}

The detailed experimental setup is shown in Fig.~\ref{A1}. The laser used in our experiment is a continuous wave intracavity frequency-doubled and frequency-stabilized Nd:YAP-LBO (Nd-doped YAlO$_{\text{3}}$ perovskite-lithium triborate) laser which generates $1080$ nm and $540$ nm light simultaneously. The $1080$ nm and $540$ nm lasers are filtered by mode cleaners (MC 1 and MC 2) respectively and then used as the seed and pump beams of the non-degenerate optical parametric amplifier (NOPA). The output mode A of the NOPA is measured by homodyne detector 1 (HD 1) while the mode B is transmitted through a lossy channel simulated by the combination of an half wave plate and a polarization beam splitter and finally detected by homodyne detector 2 (HD 2). The output signal of HD1 and HD 2 are connected to the digital storage oscilloscope for data recording.

In our experiment, the NOPA is locked with the lock-and-hold technique, which is performed with the help of two acousto-optic modulators (AOMs). The seed beam is chopped into a cyclic form with $50$ ms period, which corresponds to each locking and hold period, using two AOMs. During the locking period when AOMs are switched on, the first order of the AOM transmissions (solid red lines after two AOMs) are injected into the NOPA for the cavity locking. When the AOMs are switched off, the seed beam is chopped off (dashed red lines after two AOMs are blocked by two IDs) and the NOPA is holding. The EPR entangled state is generated and the measurement is performed during the hold period.

\section*{APPENDIX B: MANIPULATION OF THE REMOTELY SQUEEZED STATE}

The Wigner function of a pure continuous variable (CV) Einstein-Podolsky-Rosen (EPR) entangled state with correlated amplitude and anti-correlated phase quadratures is given by
\begin{equation}
\begin{aligned}
W_{AB}(x_A, p_A, x_B, p_B) =& \frac{1}{\pi^2} \exp[-\frac{(x_A - x_B)^{2}}{2e^{-2r}} - \frac{(x_A + x_B)^{2}}{2e^{2r}} \\-&\frac{(p_A - p_B)^{2}}{2e^{2r}} - \frac{(p_A + p_B)^{2}}{2e^{-2r}}],
\end{aligned}
\end{equation}
where $r$ is the squeezing parameter. The homodye measurement of quadrature with arbitrary angle $\theta$ is expressed by $\hat{x}_{\theta} = Cos\theta \hat{x} + Sin\theta \hat{p}$, where $\hat{x}$ represents the amplitude quadrature operator and $\hat{p}$ represents the phase quadrature operator, respectively. Conditioned on homodyne projective measurement on Alice's mode $x_{\theta(A)} = \alpha$, Bob's Wigner function is given by 
\begin{equation}
\begin{aligned}
W_{B}( x_B, p_B) =& \frac{1}{\pi} \exp[\alpha^{2}Sech(2r)
\\+&\frac{1}{2(1+e^{4r})}[-e^{-2r}(\alpha+Cos\theta x_{B}-Sin\theta p_{B})^2
\\-&e^{6r}(\alpha-Cos\theta x_{B}+Sin\theta p_{B})^2
\\-&2e^{2r}(\alpha^{2}+2({p_B}^{2}+{x_B}^{2})-(x_BCos\theta-p_BSin\theta)^{2})]].
\end{aligned}
\end{equation}

\begin{figure}[h]
\setcounter{figure}{0}
\renewcommand*{\thefigure}{B\arabic{figure}}
\includegraphics[width=\linewidth]{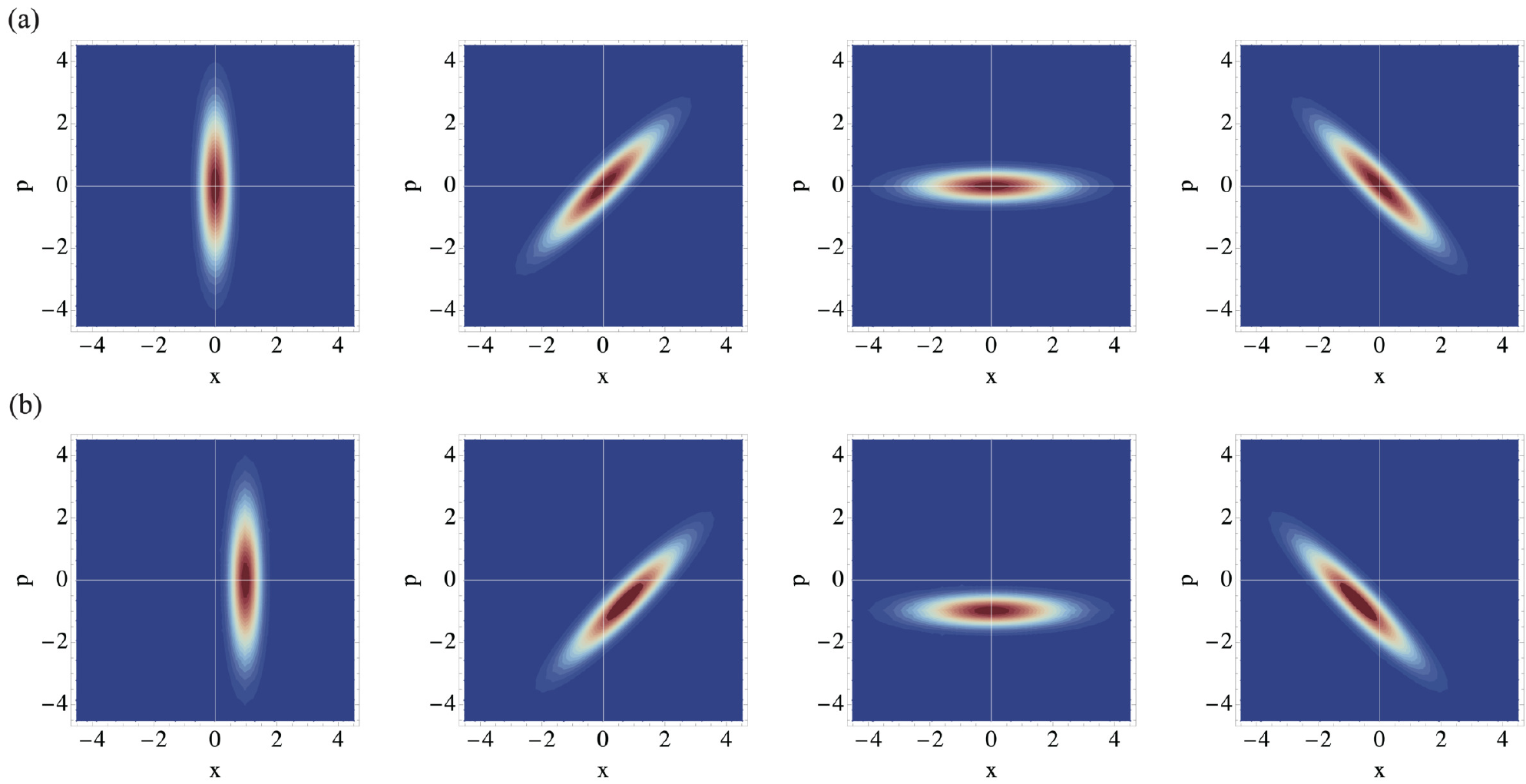}
\caption{\textbf{Contour plots of prepared squeezed states.} \textbf{(a),} Squeezed states rotated by 0, $\pi/4$, $\pi/2$ and $3\pi/4$ degrees. \textbf{(b),} Displaced squeezed states at different squeezing directions. These squeezed states have unit fidelity with pure squeezed states with $-$7.1 dB squeezing.}
\label{B1}
\end{figure}

Bob's squeezed state can be rotated by measuring Alice's quadrature operator $\hat{x}_\theta$ with different $\theta$, which is implemented by changing the relative phase between the signal and the local oscillator. For a CV EPR entangled state with $-$10 dB squeezing ($r = 1.16$), the prepared squeezed states are rotated in phase space by measuring Alice's quadrature values and projecting them on $x_{0(A)} = 0, x_{\pi/4(A)} = 0, x_{\pi/2(A)} = 0$, and $x_{3\pi/4(A)} = 0$, respectively, as shown in Fig.~\ref{B1}(a). Moreover, these squeezed states in different squeezing directions are displaced by 1 in phase space by choosing $x_{0(A)} = 1, x_{\pi/4(A)} = 1, x_{\pi/2(A)} = 1$, and $x_{3\pi/4(A)} = 1$, as shown in Fig.~\ref{B1}(b).

With the infinite squeezing of the CV EPR entangled state, homodyne projective measurement $x_{\theta(A)} = \alpha$ leads to the displacement $X = \alpha$ of the squeezed state on Bob's side. However, the value of displacement $X$ is smaller than $\alpha$ in the case of finite squeezing. In the following, we analyze the relation between the displacement and the squeezing level of the CV EPR entangled state for a fixed $\theta = 0$. For a given CV EPR entangled state shown in Eq. (5), after performing homodyne projective measurement $x_A = \alpha$ on Alice's state, the Wigner function of Bob's mode is expressed by
\begin{equation}
\begin{aligned}
W_{B}( x_B, p_B) =& \frac{1}{\pi} \exp[-Cosh(2r)(x_B- \alpha Tanh(2r))^{2} 
\\-& Sech(2r) {p_B}^{2}].
\end{aligned}
\end{equation}
It's obvious that Bob's state is an amplitude squeezed state displaced along the amplitude quadrature with displacement $X = \alpha Tanh(2r)$ and squeezing parameter of $Log[Cosh(2r)]/2$. The dependence of the squeezing level of the prepared squeezed state on the squeezing level of the CV EPR entangled state is analyzed in the main text as shown in Fig.~3(f). The dependence of the displacement $X$ on the squeezing level of the CV EPR entangled state for different projective value $\alpha$ of the amplitude quadrature is shown in Fig.~\ref{B2}. It shows that the displacement value $X$ is smaller than the homodyne projective value $\alpha$ in the case of finite squeezing and the displacement $X$ is getting close to the projective value $\alpha$ when the squeezing level is increased. 
\begin{figure}[tbp]
\renewcommand*{\thefigure}{B\arabic{figure}}
\begin{center}
\includegraphics[width=0.7\linewidth]{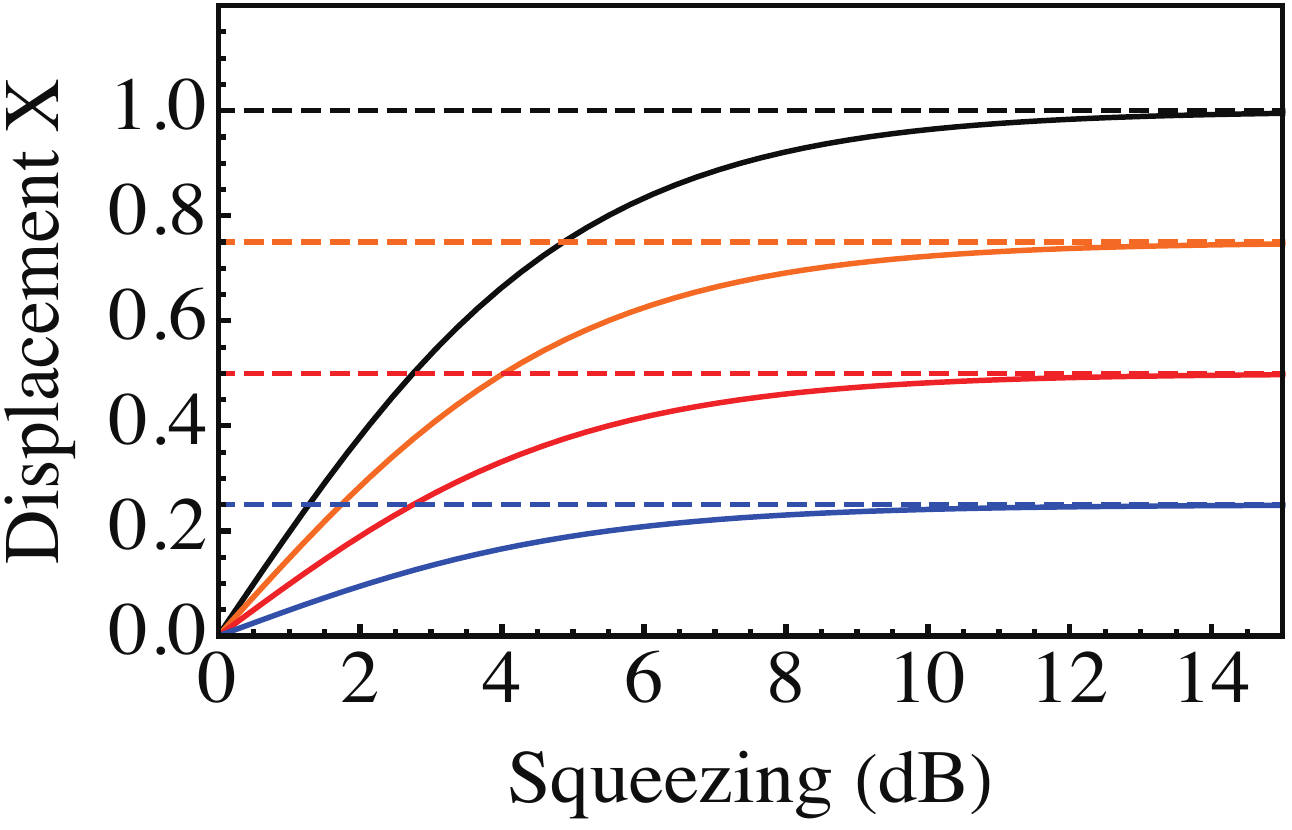}
\end{center}
\caption{\textbf{Dependence of displacements of prepared squeezed states on the squeezing level of the CV EPR entangled state.} Displacements for different values of the homodyne projective measurements $\alpha$ = 0.25, 0.5, 0.75 and 1 are shown as the blue, red, orange and black solid curves, respectively. The dotted lines represent the displacements in the case of infinite squeezing, which are equal to the projective values of homodyne measurement $\alpha$ = 0.25, 0.5, 0.75 and 1, respectively.}
\label{B2}
\end{figure}

\section*{APPENDIX C: REMOTE PREPARATION OF SQUEEZED STATE BASED ON A N-MODE CV GHZ-LIKE STATE}

The Wigner function of a $N$-mode CV Greenberger-Horne-Zeilinger-like (GHZ-like) state is given by \cite{loock2002}
\begin{equation}
\begin{aligned}
W_{N}(x, p) = &\frac{1}{\pi^{N}} \exp[-e^{-2r}[\frac{1}{N}(\sum^{N}_{i=1} x_{i})^{2}+\frac{1}{2N}\sum^{N}_{i,j} (p_{i}-p_{j})^2]
\\&-e^{2r}[\frac{1}{N}(\sum^{N}_{i=1} p_{i})^{2}+\frac{1}{2N}\sum^{N}_{i,j} (x_{i}-x_{j})^2]].
\end{aligned}
\end{equation}
It is the same to Eq. (5) when $N$ = 2. $N-1$ amplitude squeezed states can be obtained by performing the homodyne projective measurement on amplitude quadrature of an arbitrary mode, as we analyzed in the main text. Besides, because the momentum of the CV GHZ-like state is totally correlated, phase squeezed states can not be obtained by performing homodyne projective measurement only on one mode. However, by collectively performing homodyne projective measurements $p_j = 0$ on $N-1$ modes of the CV GHZ-like state, one phase squeezed state can be obtained in the rest mode. It is more clear to verify this result from the momentum wave function of the state. The momentum wave function of $N$-mode CV GHZ-like state is expressed by
\begin{equation}
\begin{aligned}
\psi_{N}(p) = (\frac{1}{\pi})^{\frac{N}{4}}\exp[-e^{2r}\frac{1}{2N}(\sum^{N}_{i=1}p_i)^2-e^{-2r}\frac{1}{4N}\sum^{N}_{i,j}(p_i-p_j)^2].
\end{aligned}
\end{equation}
After $N-1$ homodyne projective measurements $p_{j} = 0$, the momentum wave function of the rest mode $i$ becomes
 \begin{equation}
\begin{aligned}
\psi_{i}(p) = (\frac{1}{\pi})^{\frac{N}{4}}\exp[-e^{2r}\frac{1}{2N}{p_i}^2-e^{-2r}\frac{1}{2N}{p_i}^2],
\end{aligned}
\end{equation}
it is proportional to a delta function $\psi_{i}(p) \propto\delta{(p_i)}$ in the infinite squeezing which is clearly a phase squeezed state.

\begin{figure}[tbp]
\setcounter{figure}{0}
\renewcommand*{\thefigure}{C\arabic{figure}}
\begin{center}
\includegraphics[width=\linewidth]{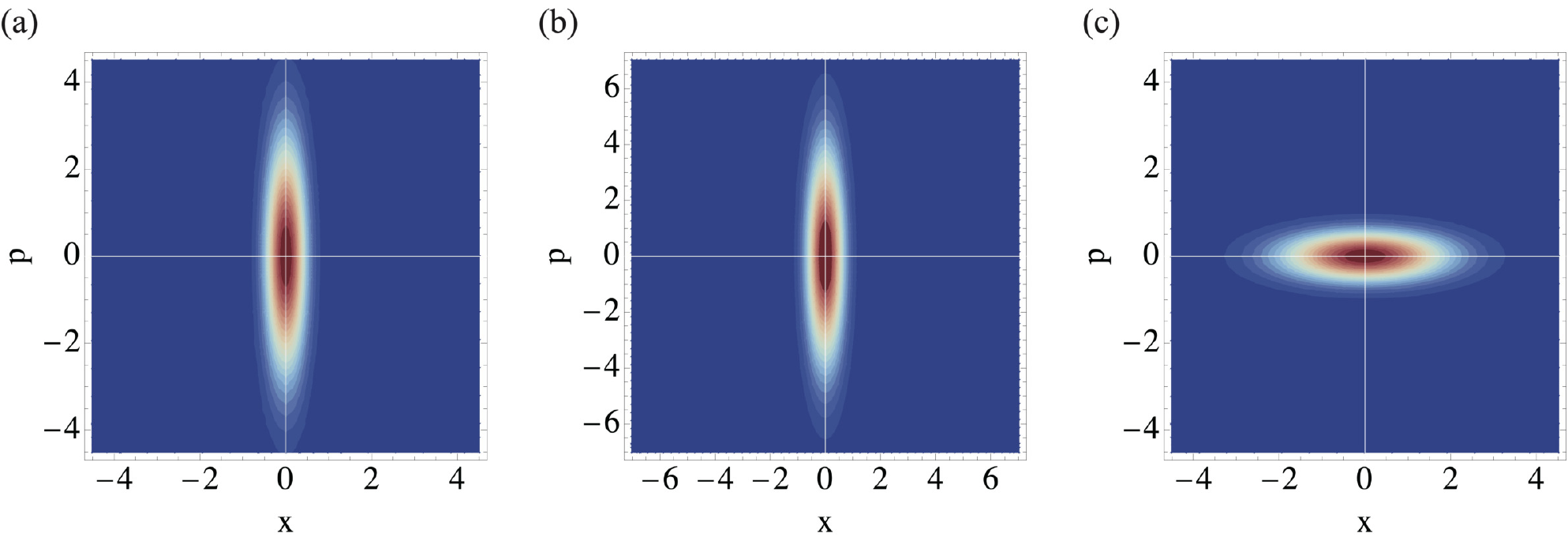}
\end{center}
\caption{\textbf{Contour plots of prepared squeezed states.} \textbf{(a,b),} Amplitude squeezed states prepared at Bob's and Claire's stations respectively by measuring Alice's mode. \textbf{(c),} Phase squeezed state prepared at Claire's station by measuring Alice's and Bob's modes.}
\label{S3}
\end{figure}

Here, we take $N = 3$ as an example. A 3-mode CV GHZ-like state is shared among Alice, Bob and Claire, the Wigner function is given by
\begin{equation}
\begin{aligned}
W_{3}(x, p) =& \frac{1}{\pi^3} \exp[-e^{-2r}[\frac{1}{3}(x_A + x_B + x_C)^{2}
\\+&\frac{1}{3}((p_A - p_B)^2 + (p_B - p_C)^2 + (p_A - p_C)^2)]
\\-& e^{2r}[\frac{1}{3}(p_A + p_B + p_C)^{2}
\\+&\frac{1}{3}((x_A - x_B)^2 + (x_B - x_C)^2 + (x_A - x_C)^2)]].
\end{aligned}
\end{equation}
By performing homodyne measurement $x_A = 0$ on Alice's mode, Wigner functions of Bob and Claire's modes become
\begin{equation}
\begin{aligned}
W_{B}(x, p) =\frac{\sqrt{3}e^{2r}\sqrt{2+e^{4r}}}{\pi+2\pi e^{4r}}\exp[\frac{-3e^{2r}}{1+2e^{4r}}{p_B}^2-\frac{e^{2r}(2+e^4r)}{1+2e^{4r}}{x_B}^2],
\end{aligned}
\end{equation}
\begin{equation}
\begin{aligned}
\\W_{C}(x, p) = \frac{\sqrt{3}e^{2r}\sqrt{2+e^{4r}}}{\pi+2\pi e^{4r}}\exp[\frac{-3e^{2r}}{1+2e^{4r}}{p_C}^2-\frac{e^{2r}(2+e^4r)}{1+2e^{4r}}{x_C}^2],
\end{aligned}
\end{equation}
respectively. Figures ~\ref{S3}(a) and ~\ref{S3}(b) show contour plots of amplitude squeezed states prepared at Bob and Claire's locations, respectively. For a 3-mode CV GHZ-like state with 10 dB squeezing, the prepared amplitude squeezed state with 7.7 dB squeezing are obtained. By performing homodyne measurements $p_A = 0$ and $p_B= 0$ on Alice and Bob's modes simultaneously, Wigner function of Claire's mode becomes
\begin{equation}
\begin{aligned}
W_{C}(x, p) = \frac{1}{\pi}\exp[\frac{2+e^{4r}}{3e^{2r}}{p_C}^2 - \frac{3e^{2r}}{2+e^{4r}}{x_C}^2].
\end{aligned}
\end{equation}
In Fig.~\ref{S3}(c), we show contour plot of a phase squeezed state prepared at Claire's location. For a 3-mode CV GHZ-like state with 10 dB squeezing, the prepared phase squeezed state with 5.4 dB squeezing and unit fidelity are obtained.

\end{document}